\documentclass[aip,reprint]{revtex4-1}
\usepackage[version=3]{mhchem}
\usepackage{graphicx}
\usepackage{miller}
\usepackage{xcolor}
\usepackage{float}
\usepackage{amssymb}

\begin{document}

\title{Sequential pulsed laser deposition of homoepitaxial \ce{SrTiO3} thin films}

\author{D.~J.~Groenendijk}
\email{d.j.groenendijk@tudelft.nl}
\affiliation{Department of Quantum Matter Physics, University of Geneva, 24 Quai Ernest-Ansermet, 1211 Gen\`eve 4, Switzerland}
\affiliation{now at Kavli Institute of Nanoscience, Delft University of Technology, Lorentzweg 1, 2628 CJ Delft, Netherlands}
\author{S.~Gariglio}
\affiliation{Department of Quantum Matter Physics, University of Geneva, 24 Quai Ernest-Ansermet, 1211 Gen\`eve 4, Switzerland}


\date{\today}

\begin{abstract}
Control of thin film stoichiometry is of primary relevance to achieve desired functionality. Pulsed laser deposition ablating from binary-oxide targets (sequential deposition) can be applied to precisely control the film composition, offsetting the importance of growth conditions on the film stoichiometry. In this work, we demonstrate that the cation stoichiometry of \ce{SrTiO3} thin films can be finely tuned by sequential deposition from \ce{SrO} and \ce{TiO2} targets. Homoepitaxial \ce{SrTiO3} films were deposited at different substrate temperatures and \ce{Ti}/\ce{Sr} pulse ratios, allowing the establishment of a growth window for stoichiometric \ce{SrTiO3}. The growth kinetics and nucleation processes were studied by reflection high-energy electron diffraction and atomic force microscopy, providing information about the growth mode and the degree of off-stoichiometry. At the optimal (stoichiometric) growth conditions, films exhibit atomically flat surfaces, whereas off-stoichiometry is accommodated by crystal defects, 3D islands and/or surface precipitates depending on the substrate temperature and the excess cation. This technique opens the way to precisely control stoichiometry and doping of oxide thin films.
\end{abstract}

\maketitle


\section{Introduction}

Transition metal oxides offer a variety of physical properties which have attracted a great deal of interest, with applications ranging from optics and electronics to sensing and actuators~\cite{Ramanathan2010}. This activity has been receiving a tremendous boost both on the theoretical side from \textit{ab-initio} calculations~\cite{Spaldin2003, Hafner2006, Seshadri2012, pentcheva2010} and on the experimental side from the ability to grow oxide thin films with single-crystal perfection~\cite{habermeier2007,schlom2008}. Among the techniques used for the growth of complex oxides, pulsed laser deposition (PLD) is the most widespread~\cite{christen2008recent, Eason2007}. For a long time, it has been considered that stoichiometric transfer from the target to the substrate could be achieved due to the high energies involved in the ablation process. However, in a set of experiments on \ce{SrTiO3} (STO) thin films~\cite{ohnishi2005,tokuda2011,breckenfeld2012,Liu2012} it was demonstrated that stoichiometric layers can only be obtained for a certain set of growth parameters such as laser fluence, spot size, and pressure. A recent study underlined the importance of the oxygen partial pressure during growth, revealing that improper plasma oxidation leads to film off-stoichiometry~\cite{orsel2015influence}. This is a particularly sensitive point as the electronic properties of complex oxides depend markedly on their stoichiometry, and any (un)intentional doping can modify their electrical and magnetic behavior. Stoichiometry control of STO thin films is important to avoid, for instance, a severe reduction of the dielectric constant~\cite{Tilley1977a,Tilley1977b,Lippmaa1999}, loss of conductivity at \ce{LaAlO3}/\ce{SrTiO3} interfaces~\cite{li2014,reinle2014}, and to obtain high electron mobilities in doped STO~\cite{kozuka2010dramatic}.

The ability to tune the composition in oxide molecular beam epitaxy (MBE) using elemental or molecular sources has been cited as a key advantage of this technique with respect to PLD~\cite{Schlom2001}. However, sequential target ablation can offer a similar stoichiometry control to PLD, offsetting the relevance of the growth conditions in determining the film composition. In this case study, we use homoepitaxial STO films to demonstrate that the cation stoichiometry in PLD-grown oxide films can be precisely tuned by sequential target ablation. STO films are deposited from \ce{SrO} and \ce{TiO2} targets with different \ce{Ti}/\ce{Sr} pulse ratios in a wide range of substrate temperatures. Using X-ray diffraction and atomic force microscopy, a growth window for stoichiometric STO is established in which the films display atomically flat surfaces. 


\section{Methods}

STO thin films were deposited on commercially available \ce{TiO2}-terminated \hkl(001)STO substrates (CrysTec GmbH) by PLD using a \ce{KrF} excimer laser (Coherent COMPexPro 205, \ce{KrF} $248\;\mathrm{nm}$). A laser energy density of 1\;$\mathrm{J}/\mathrm{cm^2}$, a laser spot size of $4\;\mathrm{mm^2}$ and a repetition rate of $1\;\mathrm{Hz}$ were used. The incident angle of the laser on the target surface was $45^\circ$ and the target-substrate distance was $5.5\;\mathrm{mm}$. The depositions were performed under an oxygen pressure of $10^{-6}\;\mathrm{Torr}$ while the chamber base pressure was $2\times10^{-9}\;\mathrm{Torr}$. In this deposition pressure, the plume propagates freely and the species arrive at the film surface with high kinetic energy, enhancing surface diffusion~\cite{xu2013impact}. The substrates were heated by an infrared laser to temperatures between $650^\circ\mathrm{C}$ and $1150^\circ\mathrm{C}$ as measured with an optical pyrometer. After growth, the samples were annealed in $0.2\;\mathrm{bar}$ \ce{O2} at $550^\circ\mathrm{C}$ for $1$ hour and cooled to room temperature in $1$ hour to refill oxygen vacancies in the films and substrates. All films were confirmed to be electrically insulating. Ceramic \ce{SrO} and single-crystal \ce{TiO2} targets were used. To control the film stoichiometry, $10$ pulses on the \ce{SrO} target were alternated with $10+n$ pulses on the \ce{TiO2} target, with $n$ ranging from $-2$ to $10$. The total number of pulses for each film was approximately 2640, corresponding to 21 u.c.~($8.2\;\mathrm{nm}$), hence the growth rate per laser shot was approximately $3\;\mathrm{pm}$ at 1 $\mathrm{J}/\mathrm{cm^2}$ laser fluence (i.e., $<1\%$ of a u.c.~per pulse). The growth was monitored by \textit{in-situ} reflection high-energy electron diffraction (RHEED). The lattice constant of the films was examined by X-ray diffraction using a PANalytical X'PertPRO MRD equipped with a monochromator. Atomic Force Microscopy (AFM) was used to measure the topography of the films. The AFM (Multimode) was operated in tapping mode with Antimony (n) doped \ce{Si} cantilevers (resonance frequency $320\;\mathrm{kHz}$, spring constant $42\;\mathrm{N}\mathrm{m}^{-1}$ and tip radius $8\;\mathrm{nm}$).


\section{Results \& Discussion}

\begin{figure}[ht!]
\includegraphics[width=\linewidth]{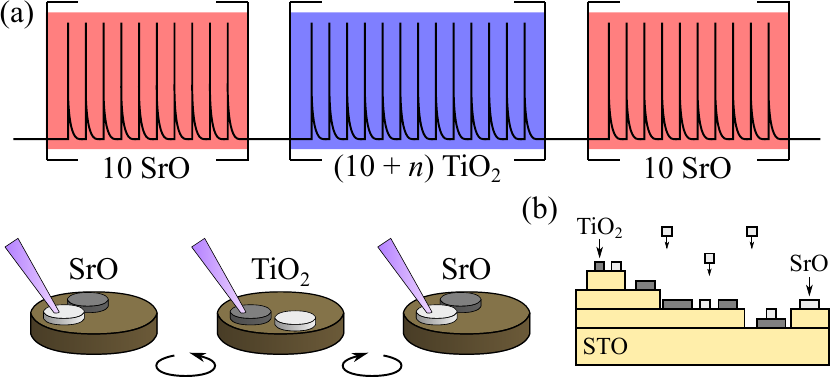}
\caption{\label{Fig1} (a) Schematic of the pulse sequence used for the growth of STO. 10 pulses of \ce{SrO} are followed by $(10 + n)$ pulses of \ce{TiO2}. The \ce{Ti}/\ce{Sr} pulse ratio is varied between 0.8 and 2 by varying $n$ from $-2$ to $10$. The multi-target carousel with \ce{SrO} and \ce{TiO2} targets rotates between deposition steps as indicated by the black arrows. (b) Schematic of the film surface during the deposition, to which sub-monolayer amounts of \ce{SrO} and \ce{TiO2} are supplied.}
\end{figure}

Figure~\ref{Fig1}(a) shows a schematic of the pulse sequence for the sequential PLD of STO thin films: \ce{SrO} and \ce{TiO2} targets, mounted on a multi-target carousel, are hit repeatedly by $10$ and $10+n$ laser pulses, respectively. In this manner, sub-monolayer amounts ($5$-$10\%$ surface coverage per cycle) of \ce{SrO} and \ce{TiO2} are supplied to the film surface where the formation of STO can occur (see Fig.~\ref{Fig1}(b)). The cation stoichiometry of the film is therefore controlled by the value of $n$.

\begin{figure}
\includegraphics[width=\linewidth]{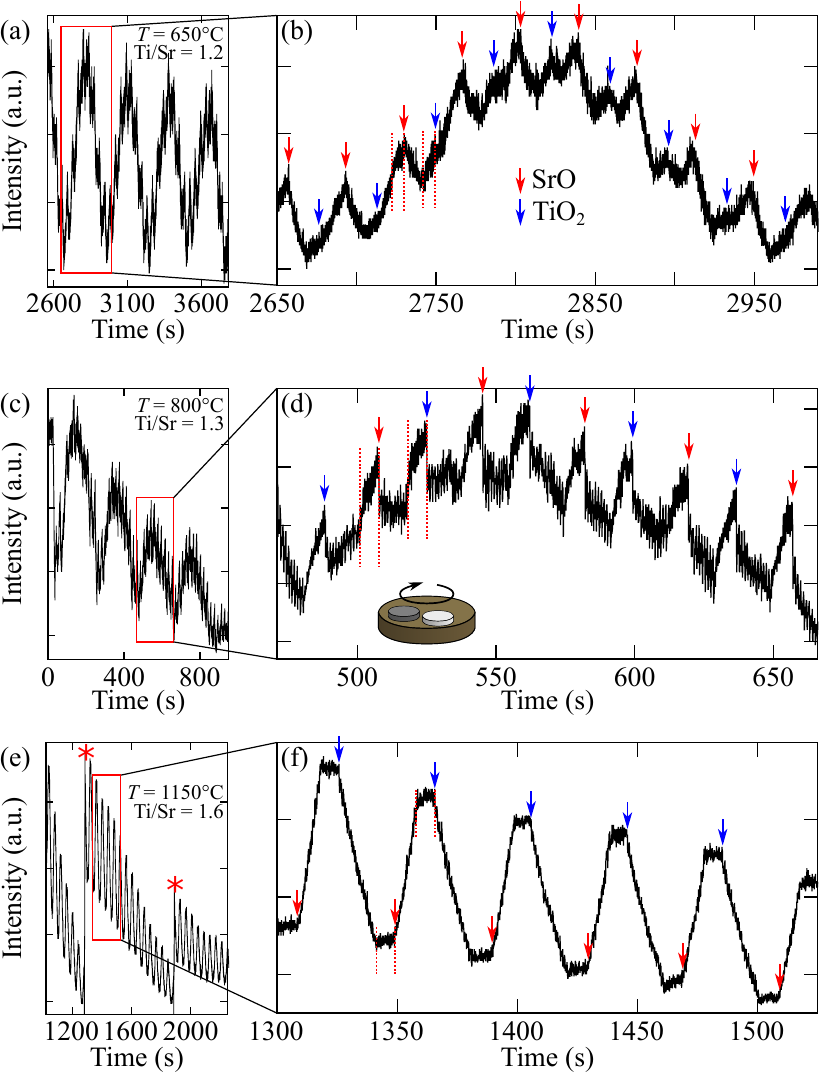}
\caption{\label{Fig2} RHEED intensity evolution of the specular spot during the sequential deposition of homoepitaxial STO thin films at $T = 650^{\circ}\mathrm{C}$ and \ce{Ti}/\ce{Sr} = 1.2 ((a) and (b)), $T = 800^{\circ}\mathrm{C}$ and \ce{Ti}/\ce{Sr} = 1.3 ((c) and (d)) and $T = 1150^{\circ}\mathrm{C}$ and \ce{Ti}/\ce{Sr} = 1.6 ((e) and (f)). The panels on the right show the intensity variation during the time required for the deposition of one unit cell of STO. The filament emission current was increased at times indicated by the red asterisks.}
\end{figure}

Previous work on PLD from a stoichiometric single-crystal STO target performed in the same pressure conditions has shown that growth occurs layer-by-layer for substrate temperatures up to $1000^{\circ}\mathrm{C}$ and in a step-flow mode at higher temperatures~\cite{Lippmaa2000}. This is a direct consequence of the temperature dependence of the surface diffusivity of the adatoms and of the supersaturation regime induced by the plume~\cite{Ferguson2009}. To gain insight into the growth kinetics in sequential deposition, we monitored the evolution of the surface morphology using \textit{in-situ} RHEED. In the in-phase diffraction condition (angle of incidence $\sim1^\circ$), the specular spot intensity can be regarded as a measure of the surface step density, reaching a minimum at the maximum step density. Figure~\ref{Fig2} shows the specular spot intensity during the deposition with optimal \ce{Ti}/\ce{Sr} pulse ratios at $T = 650^{\circ}\mathrm{C}$ ((a) and (b)), $T = 800^{\circ}\mathrm{C}$ ((c) and (d)) and $T = 1150^{\circ}\mathrm{C}$ ((e) and (f)). In general, more pulses on the \ce{TiO2} target are required for the growth of a stoichiometric film, which can be attributed to the higher reflectance of the single-crystal \ce{TiO2} target with respect to the ceramic \ce{SrO} target. The RHEED intensity recorded for a set of films deposited in a wider range of substrate temperatures and \ce{Ti}/\ce{Sr} pulse ratios is included in Fig.~S1.

At $T = 650^{\circ}\mathrm{C}$ (Fig.~\ref{Fig2}(a)) and $T = 800^{\circ}\mathrm{C}$ (Fig.~\ref{Fig2}(c)), clear oscillations of the intensity envelope corresponding to the growth of single STO unit cells are observed, indicating a layer-by-layer growth mode. We confirmed that the number of oscillations yields a layer thickness in agreement with the estimation from X-ray diffraction by studying a film deposited on a \ce{DyScO3} substrate (see Fig.~S2). At $T = 1150^{\circ}\mathrm{C}$ (Fig.~\ref{Fig2}(e)), such intensity oscillations are not observed: at this temperature the growth occurs in step-flow mode, where the overall intensity remains constant but shows a strong increase or decrease during the deposition of each atomic specie~\footnote{The overall decrease in intensity is mainly caused by a decline of the filament emission current, which was manually increased at times indicated by the red asterisks.}. Figure~\ref{Fig2}(b), (d) and (f) show the RHEED intensity recorded over a time period required for the deposition of one unit cell: here, the \ce{SrO} and \ce{TiO2} deposition steps as well as the rotation of the multi-target carousel (switching time $\sim 7\;\mathrm{s}$) are visible.

In low-temperature, single-target PLD growth, due to the low surface diffusivity, intra- and interlayer transport occurring during the thermalization of the laser plume on the sample surface dominates over--it is at least three orders of magnitude faster than--thermal equilibrium transport: growth proceeds by island nucleation and coarsening~\cite{Tischler2006}. Figures~\ref{Fig2}(b) and \ref{Fig2}(d) show that this is also the case  for sequential deposition at optimal \ce{Ti}/\ce{Sr} pulse ratios. Both at $650^\circ\mathrm{C}$ and $800^\circ\mathrm{C}$, a strong, exponential recovery of the intensity is observed during the target switching time. During this time, the surface step density is reduced by smoothening of 2D island contours, island coalescence and formation of STO from \ce{SrO} and \ce{TiO2}. At $800^\circ\mathrm{C}$, the large intensity drop directly after each laser pulse point towards an atomically smooth surface prior to the deposition, suggesting that the surface diffusivity is sufficiently high for the aforementioned processes to occur during the target switching time. We notice that the presence and persistence of the intensity oscillations, usually a sign of the preservation of the surface smoothness~\cite{Eres2002}, are also a clear indication of the correct stoichiometry: moving to higher or lower \ce{Ti}/\ce{Sr} pulse ratios results in a loss of RHEED intensity after a few unit cells (see Fig.~S1), indicating a strong roughening of the surface. In near-stoichiometric conditions, the oscillations of the intensity envelope disappear after a few unit cells (Fig.~S4) while the RHEED pattern remains 2D, which we attribute to a crossover from layer-by-layer to step-flow growth. This has previously been observed for films deposited from a single STO target under stoichiometric conditions~\cite{xu2013impact}. We will show later on that this crossover can be related to the evolution of the surface morphology during growth. 

At $T = 1150^{\circ}\mathrm{C}$ (Fig.~\ref{Fig2}(f)), no oscillation envelope is observed and the intensity sharply increases (decreases) after every pulse of the \ce{SrO} (\ce{TiO2}) deposition step, whereas it remains constant during the switching time. The large amplitude of the RHEED intensity variation is a sign of an atomically smooth surface prior to deposition. Together these observations point towards a large diffusion length of \ce{Sr} and \ce{Ti} and a short time scale for the formation of STO. The absence of intensity oscillations and the full recovery after the two-target deposition sequence indicate that the growth proceeds in a step-flow mode. Such behavior is also observed after a few unit cells during the deposition at $800^\circ\mathrm{C}$ in near-stoichiometric conditions. This is different from the single-target deposition of STO at high temperature where the intensity remains constant as it fully recovers on a short time scale ($<1\;\mathrm{s}$) after each laser pulse~\cite{Lippmaa2000}. It is instead reminiscent of the intensity evolution reported during the growth of STO by MBE using shuttered deposition~\cite{Bodin1992,Iijima1992,Haeni2000}, and by sequential PLD of complete monolayers of \ce{SrO} and \ce{TiO2}~\cite{herklotz2015stoichiometry}. In these cases, it is also the deposition of each atomic specie (\ce{Sr} or \ce{Ti}) that drives the decrease/increase of the diffracted intensity. Such behavior is however reported for much lower deposition temperatures (550-750$^{\circ}\mathrm{C}$) where the growth proceeds by alternating complete monolayers of \ce{SrO}/\ce{TiO2} in a layer-by-layer fashion. Here, the evolution of the intensity is found to be related to the stoichiometry as shown in Fig~S3. For low \ce{Ti}/\ce{Sr} pulse ratios, the intensity decreases during the \ce{SrO} deposition, whereas at higher pulse ratios it decreases during the \ce{TiO2} deposition. This indicates that the deposition of the excess cation increases the surface step density. 

\begin{figure}[ht!]
\includegraphics[width=\linewidth]{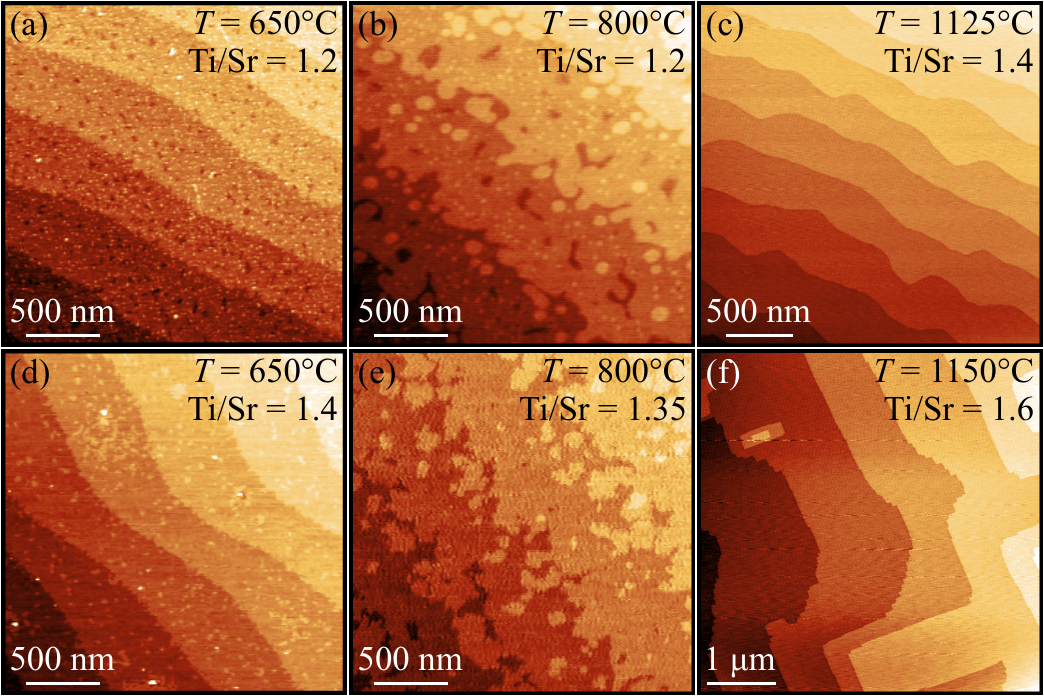}
\caption{\label{Fig3} AFM topographic images of homoepitaxial STO films grown at different temperatures and \ce{Ti}/\ce{Sr} pulse ratios. (a) $T = 650^{\circ}\mathrm{C}$, $\ce{Ti}/\ce{Sr} = 1.2$. (b) $T = 800^{\circ}\mathrm{C}$, $\ce{Ti}/\ce{Sr} = 1.2$. (c) $T = 1125^{\circ}\mathrm{C}$, $\ce{Ti}/\ce{Sr} = 1.4$. (d) $T = 650^{\circ}\mathrm{C}$, $\ce{Ti}/\ce{Sr} = 1.4$. (e) $T = 800^{\circ}\mathrm{C}$, $\ce{Ti}/\ce{Sr} = 1.35$.  (f) $T = 1150^{\circ}\mathrm{C}$, $\ce{Ti}/\ce{Sr} = 1.6$.}
\end{figure}

The topography of the homoepitaxial STO films, studied after deposition by AFM, confirms the observed growth modes. Figure~\ref{Fig3} shows AFM topographic images of films deposited at $T = 650^{\circ}\mathrm{C}$ ((a) and (d)), $T = 800^{\circ}\mathrm{C}$ ((b) and (e)) and $T = 1125$-$1150^{\circ}\mathrm{C}$ ((c) and (f)). The step heights for all films are $\sim4\;\mathrm{\AA}$, showing that the surface is fully \ce{SrO}- or \ce{TiO2}-terminated. 

The surface of the STO film grown at $T = 650^{\circ}\mathrm{C}$ shows small 2D islands, which is coherent with the small adatom mobility and the large nucleation at this temperature. In addition, the step edge of the substrate is not modified by the growth. This changes upon increasing the substrate temperature: at $T = 800^{\circ}\mathrm{C}$, the film surface displays flat terraces but the step edges become roughened due to the proximity/aggregation of unit-cell high islands. This is consistent with the strong intensity recovery during the target switching time in Fig.~\ref{Fig2}(d), in which larger islands grow at the expense of smaller ones, reducing the step density, and the island contours smoothen. Not only the temperature, but also the cation stoichiometry is found to affect the surface morphology of the films. At $800^\circ\mathrm{C}$ and \ce{Ti}/\ce{Sr} = 1.2, holes and 2D islands are simultaneously present on the terraces, indicating that the growth occurs in a two-layer growth mode and that there is significant interlayer transport. When the \ce{Ti}/\ce{Sr} ratio is increased, the density of holes on the terraces decreases, and at \ce{Ti}/\ce{Sr} = 1.35 (Fig.~\ref{Fig3}(e)) the island shape changes from circular to ``dendritic'' or ``fractal-like''. Such island shapes can be advantageous for obtaining ideal 2D layer-by-layer or step-flow growth, since the average distance from any point on top of an island to an edge site is smaller than in the case of circular islands~\cite{christen2008recent}. As mentioned above, the RHEED intensity variation during the deposition of near-stoichiometric films at $800^\circ\mathrm{C}$ showed that the growth mode evolved from layer-by-layer to step-flow. Such a crossover can occur when the diffusion length is smaller than the terrace width and therefore initially not large enough to obtain a step-flow mode. This is in agreement with the AFM images shown in Fig.~\ref{Fig3}(b) and (e), where the step edges are roughened and the effective terrace length is small enough for 2D islands to diffuse to the terrace edges. 

At higher temperatures ($T = 1125$-$1150^{\circ}\mathrm{C}$), no 2D islands are observed and the step edges are continuous, confirming the high adatom mobility and the resulting step-flow growth mode. The entire set of AFM topographic images of the films deposited at different substrate temperatures and \ce{Ti}/\ce{Sr} pulse ratios is included in the supplementary material (Fig.~S5). Similar to previous reports, excess \ce{Sr} results in 3D island growth after several layers, while excess \ce{Ti} still yields a flat surface~\cite{xu2013impact}. This can be due to the lower solubility limit of \ce{Ti}-vacancies than \ce{Sr}-vacancies, where the former can result in stacking faults~\cite{keeble2013nonstoichiometry}. In addition, excess \ce{Ti} could be accommodated by migration of \ce{Sr} from the STO substrate as has previously been observed for \ce{TiO2} epitaxy on STO~\cite{radovic2011situ}.

We now turn our attention to the study of the film stoichiometry by X-ray diffraction. It is well established that the lattice constant of STO films can be used as a sensitive indicator of cation off-stoichiometry~\cite{ohnishi2008,brooks2009,jalan2009}. First-principle calculations show that the formation of \ce{Sr} vacancies is energetically favorable over the formation of \ce{Ti} vacancies~\cite{tanaka2003}. However, in both \ce{Sr}- and \ce{Ti}-excess STO thin films, a lattice expansion is observed. On the \ce{Ti}-rich side, \ce{Sr} vacancies or the inclusion of \ce{TiO2} planes forming a Magn\'eli phase~\cite{li2013} induce an increase of the $c$-axis. In a detailed TEM study, this has been attributed to the clustering of \ce{Sr} vacancies which expand the lattice by their Coulombic repulsion~\cite{tokuda2011}. On the \ce{Sr}-rich side, \ce{O} and \ce{Sr} form a Ruddlesden-Popper (RP) phase \ce{Sr_{n+1}Ti_{n}O_{3n+1}}, which determines an increase of the $c$-axis.

\begin{figure}
\includegraphics[width=\linewidth]{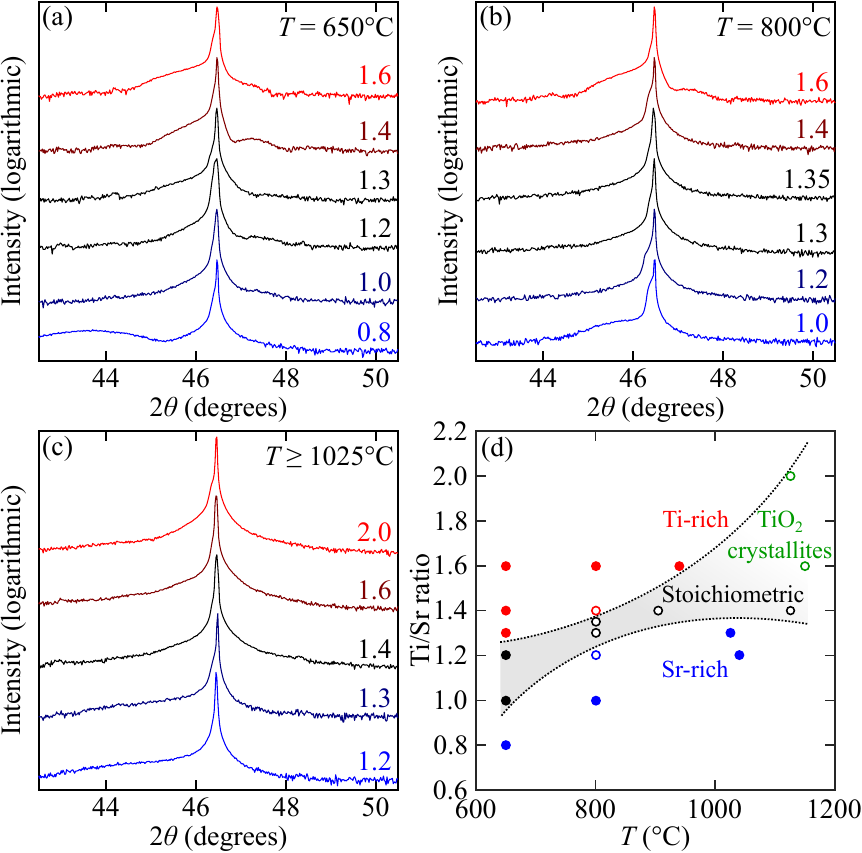}
\caption{\label{Fig4} $2\theta$-$\omega$ X-ray diffraction measurements around the \hkl(002) peak of homoepitaxial STO films with different \ce{Ti}/\ce{Sr} pulse ratios deposited at (a) $T = 650^\circ\mathrm{C}$, (b) $T = 800^\circ\mathrm{C}$, and (c) $T > 1025^\circ\mathrm{C}$. (d) Growth window for stoichiometric STO films with atomically flat surfaces. The open circles correspond to films for which a step-flow growth mode is obtained.} 
\end{figure}

In homoepitaxial STO films, one can therefore consider the \hkl(002) peak of the substrate and use the deviation of the film peak as a measure of the cation off-stoichiometry. Figure~\ref{Fig4} shows $2\theta$-$\omega$ X-ray diffraction measurements in the range $2\theta = 42.5^\circ$ to $50.5^\circ$ with \ce{Ti}/\ce{Sr} pulse ratios from $0.8$ to $2.0$ grown at $T = 650^\circ\mathrm{C}$ (a), $T = 800^\circ\mathrm{C}$ (b) and $T > 1025^\circ\mathrm{C}$ (c). It is sufficient to consider this range as measurements over a larger range show that no additional phases are formed (Fig.~S6).

At $T = 650^\circ\mathrm{C}$ (Fig.~\ref{Fig4}(a)) and $T = 800^\circ\mathrm{C}$ (Fig.~\ref{Fig4}(b)), a peak is observed to the left of the substrate peak at the lowest \ce{Ti}/\ce{Sr} pulse ratio, indicating films with an enlarged $c$-axis. As the \ce{Ti}/\ce{Sr} pulse ratio is increased, this peak moves closer to the substrate peak, overlaps, and moves away for $\ce{Ti}/\ce{Sr} \geq 1.4$. This indicates that the film stoichiometry can be tuned from \ce{Sr}-rich to \ce{Ti}-rich, enabling the establishment of a growth window for stoichiometric STO as shown in Fig.~\ref{Fig4}(d). At the highest growth temperatures (Fig~\ref{Fig4}(c)), increasing the \ce{Ti}/\ce{Sr} ratio reduces the $c$-axis, suggesting a transition from \ce{Sr}-rich to stoichiometric films. In general, increasing the substrate temperature is found to widen the range of pulse ratios in which stoichiometric films are obtained (see Fig.~S7), which is consistent with the persistence of intensity oscillations and the formation of atomically flat surfaces. Similar to what was observed previously with MBE~\cite{jalan2009}, the optimal \ce{Ti}/\ce{Sr} pulse ratio increases with increasing temperature, which could be related to higher volatility of one of the deposited species or compensation of \ce{Sr}-deficiency by migration of \ce{Sr} from the STO substrate. Not only the pulse ratio, but also the number of pulses per cycle plays a role in determining the film quality. Figure~S8 shows that increasing the number of pulses per cycle while keeping the ratio fixed results in lower film quality, as evidenced by a loss of RHEED intensity, roughening of the film surface and an enlarged $c$-axis. This could be due to the larger amount of material supplied per cycle, which can result in clustering of \ce{SrO} and/or \ce{TiO2} as it does not allow enough time for the formation of STO. We note that this behavior differs from the report of Herklotz et al.~\cite{herklotz2015stoichiometry}. This may be due to the different oxygen pressures used in the two studies: in our growth conditions ($10^{-6}\;\mathrm{Torr}$), the plume propagates freely until reaching the substrate while in ~\cite{herklotz2015stoichiometry} the oxygen gas ($0.02\;\mathrm{Torr}$) stops the plume expansion, thermalizing the species before their arrival at the film surface~\cite{amoruso2006propagation}. 

\begin{figure}
\includegraphics[width=\linewidth]{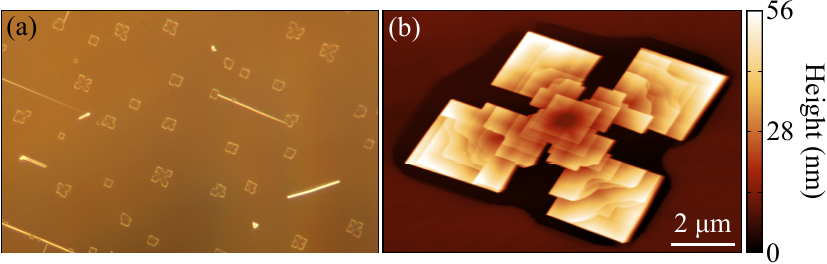}
\caption{\label{Fig5} (a) Dark-field optical microscopy image showing micron-sized rod- and flower-shaped \ce{TiO2} crystallites formed on the STO surface. The STO deposition conditions were $1120^\circ\mathrm{C}$ and $\ce{Ti}/\ce{Sr} = 2.0$. (b) AFM topographic image of a single, flower-shaped crystallite.} 
\end{figure}

At the highest growth temperatures, films with $\ce{Ti}/\ce{Sr} = 1.2$ and $\ce{Ti}/\ce{Sr} = 1.3$ display a broad peak to the left of the substrate peak (Fig.~\ref{Fig4}(c)), which disappears when the \ce{Ti}/\ce{Sr} pulse ratio is increased. This peak is most likely related to the presence of excess \ce{SrO}, which migrates to the surface and form islands as appear in the AFM topographic image in Fig.~S5. On the other hand, films deposited with rather high \ce{Ti}/\ce{Sr} pulse ratios ($1.4$ to $2.0$) appear to be stoichiometric. In fact, at these temperatures the STO film is found to accommodate excess \ce{Ti} in crystallites on the surface. Figure~\ref{Fig5}(a) shows a dark-field optical microscopy image of the surface of a \ce{Ti}-excess STO film deposited at high temperature ($T = 1120^\circ\mathrm{C}, \ce{Ti}/\ce{Sr} = 2.0$). Rod-, square- and flower-shaped crystallites with micrometer dimensions are sufficiently large to be observed by optical microscopy. The square- and flower-shaped crystallites form along the principal axes of the substrate, whereas the rod-shaped crystallites additionally appear at $45^\circ$ angles. In Fig.~\ref{Fig5}(a), different stages of the growth of the crystallites can be identified: small, square-shaped crystallites tend to grow larger in size before branching. An STO film with lower \ce{Ti}/\ce{Sr} pulse ratio ($\ce{Ti}/\ce{Sr} = 1.6$) deposited at $1150^\circ\mathrm{C}$ contains crystallites of smaller density and size, corroborating that the formation is driven by excess \ce{Ti} (Fig.~S9(b)). An AFM topographic image of a flower-shaped crystallite is shown in Fig.~\ref{Fig5}(b). It has a height of approximately $30\;\mathrm{nm}$ and is faceted down towards the center. The edges are parallel to the in-plane principal axes of the STO substrate. This suggests outward growth, where \ce{Ti} adatoms arriving within a distance in the order of the diffusion length can attach to the outer edges of the crystallite. The area surrounding the crystallite is approximately $8\;\mathrm{nm}$ below the rest of the film, which is the same value as the expected film thickness, indicating that the growth of this crystallite commenced as soon as the deposition started. Hence, at high temperature, the formation of \ce{TiO_x} crystallites appears to be an energetically favorable way to accommodate excess \ce{Ti}.


\section{Conclusions}

In summary, we showed that the cation stoichiometry of homoepitaxial STO thin films can be finely tuned by sequential PLD from \ce{SrO} and \ce{TiO2} targets. The growth kinetics, monitored by \textit{in-situ} RHEED, display a transition from layer-by-layer to step-flow growth at higher temperatures. The surface morphology of the films was studied by AFM, confirming the growth modes observed by RHEED. X-ray diffraction measurements showed that the film stoichiometry can be tuned from \ce{Sr}-rich to \ce{Ti}-rich by increasing the \ce{Ti}/\ce{Sr} pulse ratio. At high temperature $(>1120^\circ\mathrm{C})$ and high \ce{Ti}/\ce{Sr} pulse ratios, excess \ce{Ti} was found to be accommodated in crystallites on the STO surface rather than incorporation of point defects or re-evaporation. At the optimal growth conditions, stoichiometric thin films with atomically smooth surfaces were obtained. This work provides insight into the growth kinetics of homoepitaxial STO thin films from binary-oxide targets and opens a promising route towards precise stoichiometry control and doping of oxide thin films. The technique can be particularly useful for the synthesis of materials that are not stable in bulk form or when growth is complicated by incongruent ablation or preferential scattering of lighter species.


\begin{acknowledgments}
The authors thank Danfeng Li, Jennifer Fowlie and Jean-Marc Triscone for valuable discussions. This research was supported by the Swiss National Science Foundation and the MaNEP association.\\
\end{acknowledgments}


\bibliographystyle{apsrev4-1}
\bibliography{References}

\end{document}